\documentclass[aps,pra,paper,preprint,amsmath,amssymb,groupedaddress,floatfix,showkeys,subeqn]{revtex4-1} 

\usepackage{hyperref}
\usepackage{graphics} 
\usepackage{graphicx} 
\usepackage{subfigure}
\usepackage{url}
\usepackage{tipa}

\begin{document}

\title{Multistate APLIP and VibLIP:\\ From molecular bond extension to atomic transport}

\author{Kalle-Antti Suominen}
\affiliation{Turku Centre for Quantum Physics, Department of Physics and Astronomy, University of Turku, FIN-20014 Turku, Finland}

\begin{abstract}
APLIP is a method for using a STIRAP-like three-state configuration with two laser pulses for continuous extension of a molecular bond, introduced in 1998. It is based on time-dependent light-induced potential surfaces (LIP). In VibLIP one extends the idea into a method for tailoring the vibrational state while changing the electronic state, introduced in 2000. Here I discuss the extension of both methods to situations that involve more than three electronic states, and note the possibility of using the method on adiabatic transport of atoms between microtraps or equivalent structures. 
\end{abstract}
\bigskip

\keywords{Molecular bond extension, Atomic transport, STIRAP}

\maketitle

\section{Introduction}

\begin{figure}
\begin{center}
\includegraphics[width=16cm]{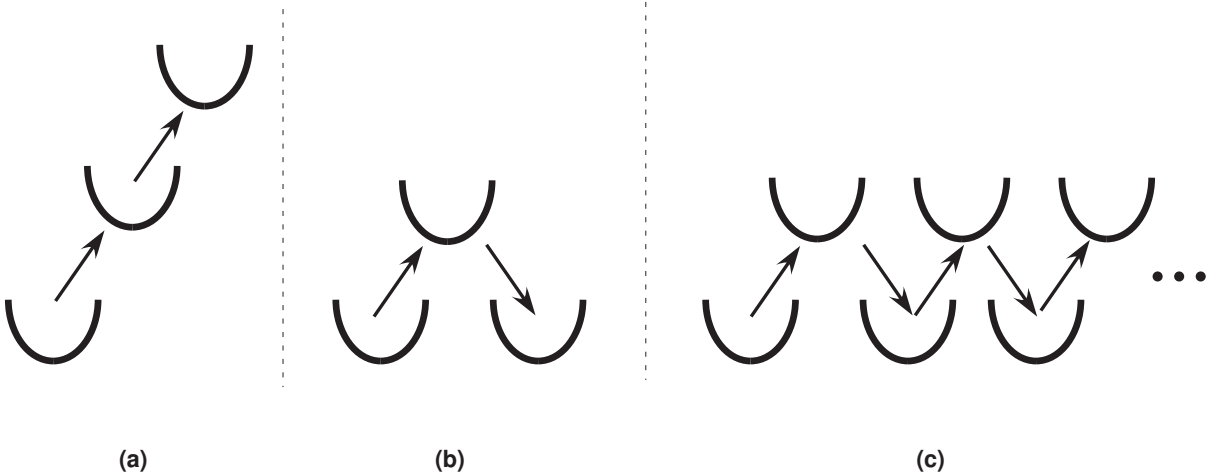}
\caption{APLIP schemes between states.}
\label{MultiAPLIP}
\end{center}
\end{figure}

Manipulation of dimer molecules with strong laser pulses holds strong interest both in physics and in chemistry~\cite{Garraway1995,Garraway2002,May2011}. The toolbox for such manipulation is nowadays quite extensive. One special topic is the control of molecular bond length, mainly the extension of the bond, see e.g.~\cite{Sola2000,Sola2004,Chang2010,Chang2011,Chang2012}. One can achieve that simply via efficient change of the electronic molecular state, going from one vibrational ground state to another. Optimization of the process requires robustness, efficiency and minimization of vibrational excitation. In 1998 Garraway and Suominen introduced APLIP~\cite{Garraway1998}, i.e. Adiabatic Passage on Light-Induced Potentials. This method for bond extension is related to STIRAP as it also relies on counterintuitive pulse sequence and minimal excitation of the intermediate state in the three-state process. However, unlike STIRAP, it basically ignores the vibrational state structure and all dynamics is based on the time-dependent eigenstates that arise when the Hamiltonian with the electronic molecular states (Born-Oppenheimer states) and laser-induced couplings is diagonalised. These states are the Light-Induced Potentials (LIP) that depend both on time as well on the nuclear coordinate. The method is quite robust as discussed in ref.~\cite{Garraway2003}. The process is sketched in fig.~\ref{MultiAPLIP}(a). The basic idea is that the pulses couple the three levels sequentially but in counterintuitive order, and they are quite off-resonant in the direct transitions, but achieve the two-photon resonance between the initial and the final state.

Interestingly, although the method relies on adiabaticity, it is clear with afterthought that it can not be completely adiabatic, since then the final state should equal the initial state. In fact, when looking only at light-induced potentials we easily miss the fact that when the pulses are weak and the original (diabatic) electronic states dominate, the system undergoes transitions between LIP states by simply remaining on the original states. This aspect is studied in quite detail in refs.~\cite{Rodriguez2000,Harkonen2006}. By controlling these diabatic transitions we can develop APLIP further into a tool for combining the change of the electronic state with a highly selective excitation of a vibrational state; this was discovered by Rodriguez et al. in 2000~\cite{Rodriguez2000}. For simplicity we use the acronym VibLIP for the process. It opens a new possibility for control of vibrational states, but it lacks the robustness of APLIP~\cite{Rodriguez2000,Harkonen2006}. 

The use of strong pulses is problematic for molecular applications since one can expect multiphoton processes to affect the process~\cite{Garraway1998}. The control of molecular bond, however, has many similarities with the control of atomic motion. The simplest mapping would be to replace the nuclear coordinate with the atomic position, the electronic state potentials with traps achieved with external fields (magnetic or laser fields) and the couplings are produced with lasers or with magnetic rf-fields or microwave fields. The molecular electronic states would map to atomic electronic states and their spin-substates. Note that in APLIP the intermediate state does not need to have bound states; it can be a dissociative state or even an inverted parabolic potential (which can occur in cold atom physics). Also, both APLIP and VibLIP do not require the ladder structure of fig.~\ref{MultiAPLIP}(a) but work also for the lambda-case of fig.~\ref{MultiAPLIP}(b), which is often the more appropriate case in atomic physics. 

The purpose of this work is not to discuss the specific cases of such mappings. Two important questions remain when considering the world of cold atoms. One is the applicability of APLIP and VibLIP beyond the one-dimensional treatments of refs.~\cite{Garraway1998,Rodriguez2000,Garraway2003}. This has been studied by H\"ark\"onen et al.~\cite{Harkonen2006}, and it has been shown that the approach can be extended into two dimensions. The second aspect, studied in this manuscript, is how well the two methods apply to systems that have more than three sequentially coupled states, such as fig.~\ref{MultiAPLIP}(c). In the following I present that both methods can be extended to systems with a higher number of states (assuming, though, that the number is always odd) and I discuss the details that arise in such extensions.   

\section{Multistate APLIP and VibLIP}

The Hamiltonian for the three-state APLIP and VibLIP processes is simply~\cite{Garraway1998,Rodriguez2000,Harkonen2006}
\begin{equation}
   H = -\frac{\hbar^2}{2m}  \frac{ \partial^2 }{ \partial R^2 }\
       {\cal I} + {\cal U}(R,t)
   \label{ham}
\end{equation}
where $R$ is the nuclear separation coordinate, $m$ is the reduced mass of the 
molecule, and the electronic potentials and laser-induced couplings are given by
\begin{equation}
   {\cal U}(R,t) = \left[
   \begin{array}{ccc}
      U_1(R)+\hbar\Delta_1 & \hbar\Omega_A(t)  & 0 \\
      \hbar\Omega_A(t)  & U_2(R)+\hbar\Delta_2 & \hbar\Omega_B(t) \\
      0 & \hbar\Omega_B(t)   & U_3(R)+\hbar\Delta_3
   \end{array} \right].
   \label{U}
\end{equation}
This description requires, though, that the rotating wave approximation (RWA) is applicable~\cite{Garraway1995}. The extension of this tri-diagonal matrix into more than three states is trivial. In the original APLIP study~\cite{Garraway1998} realistic potentials and units for a sodium dimer were used, but here we go for scaled units ($\hbar=1$, $m=1/2$, and denote distance with $x$) and for simple harmonic potentials and Gaussian pulses:
\begin{equation}
  U_i(x)=\Delta_i+A_i(x-x_i)^2,\quad i=1,2,...\qquad \Omega_j(t) = B_j\exp[-(t-t_j)^2/(\Delta t_j)^2],
  \quad j=A,B,...
\end{equation}

\begin{figure}
\begin{center}
\begin{minipage}{160mm}
\subfigure[]{
\resizebox*{8cm}{!}{\includegraphics{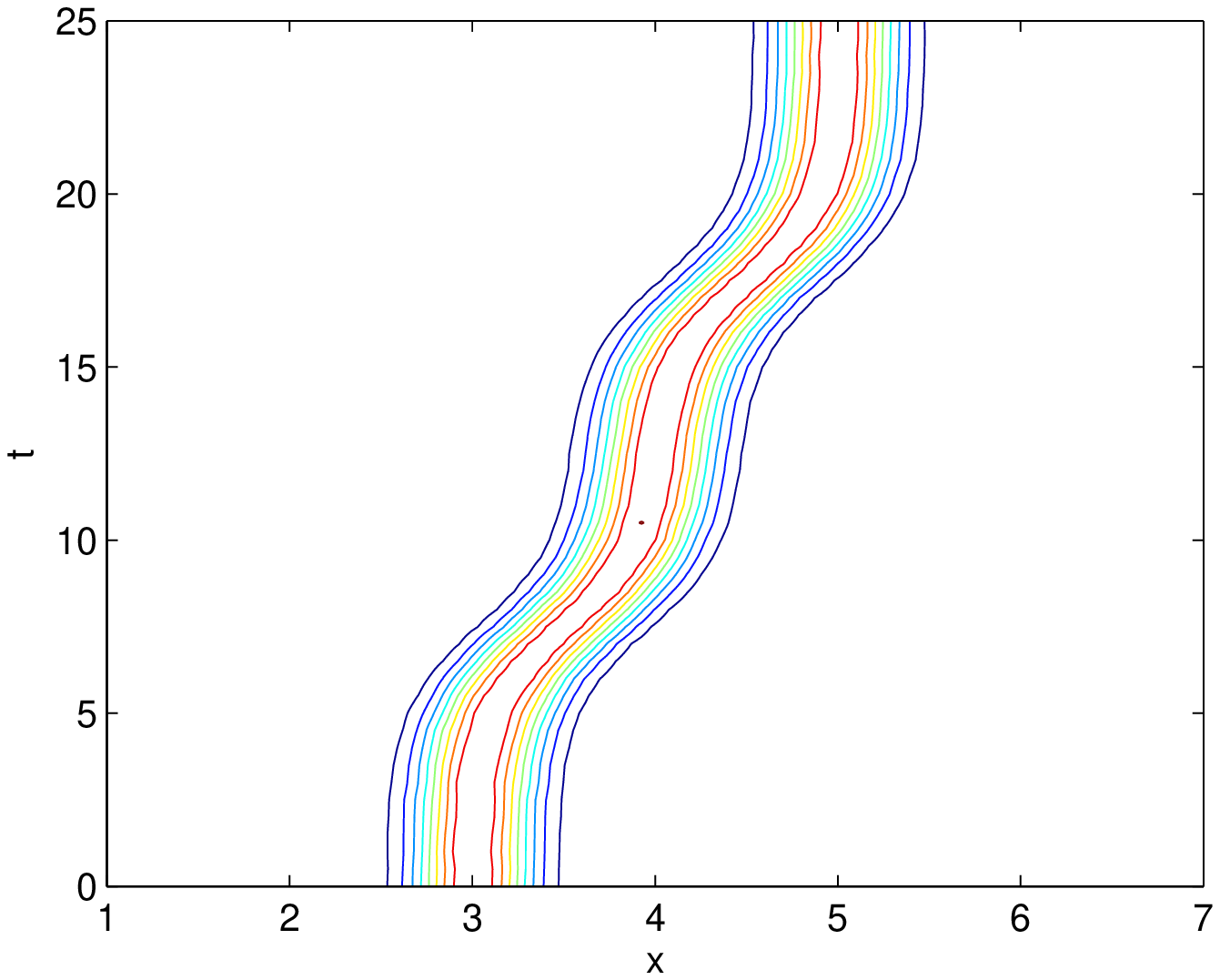}}}%
\subfigure[]{
\resizebox*{8cm}{!}{\includegraphics{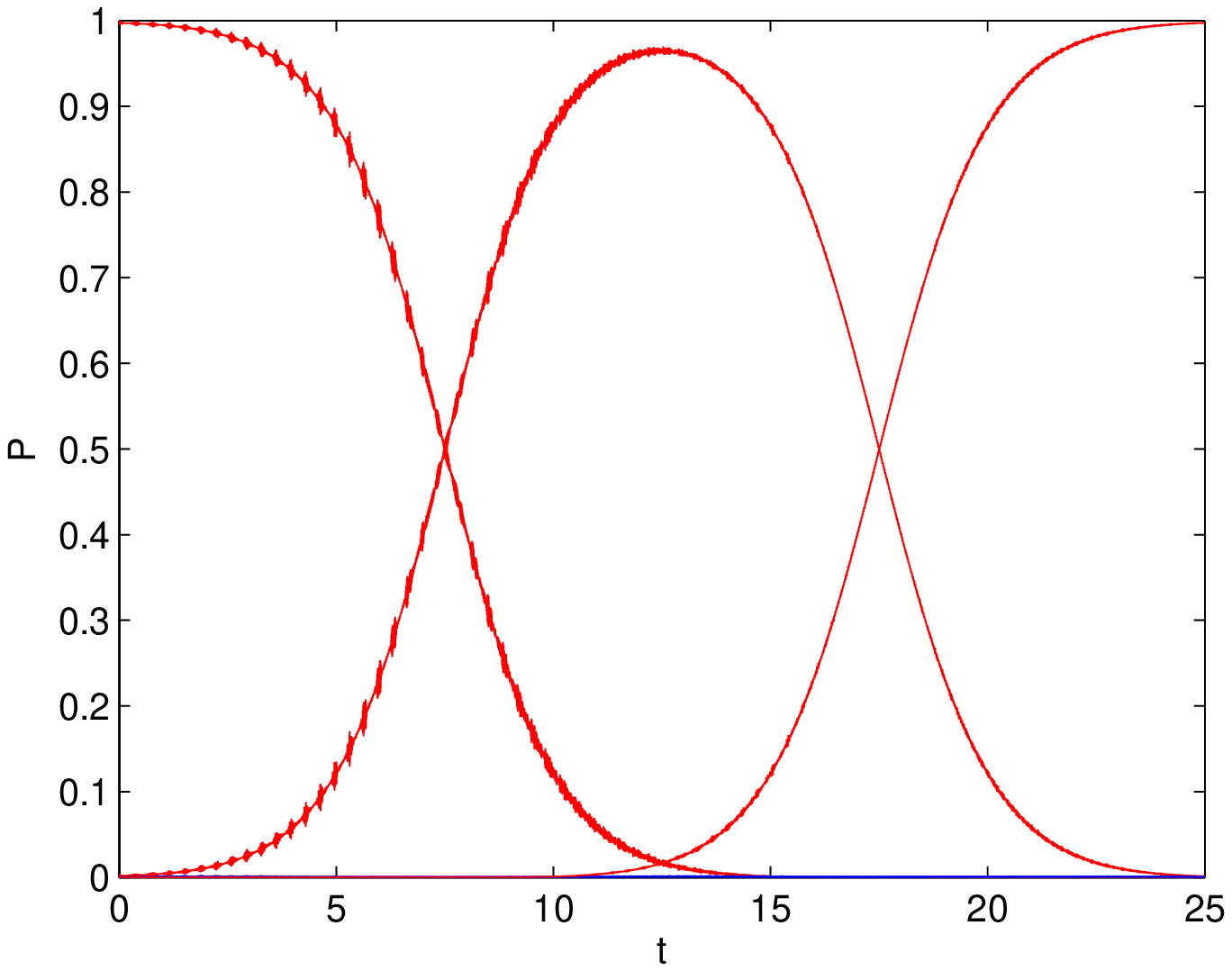}}}%
\caption{APLIP process with five states as two sequential three-state APLIP processes.}%
\label{DoubleThree}
\end{minipage}
\end{center}
\end{figure}

Let us start with the simplest extension, which is a five-state system with four pulses. An obvious extension is to have just two separate three-state APLIP processes as a sequence, which means that one of the intermediate states (nr. 3) is fully occupied during the process. In fig.~\ref{DoubleThree} we show such a process. The contour plot (a) shows how the Gaussian ground state of the harmonic potential moves to higher $x$ as a stepwise process, and the state populations in (b) show how the middle state (nr. 3) is almost fully populated, but the two other intermediate states (nr. 2 and 4) remain unoccupied. Here the scaled parameters are $\Delta_1=\Delta_3=\Delta_5=0$, $\Delta_2=\Delta_4=-300$, $A_i=100$ for $i=1,..,5$, $\Delta t_j=5,B_j=500$ for $j=A,B,C,D$, $x_1=3.0,x_2=3.5,x_3=4.0,x_4=4.5,x_5=5.0$, and $t_A=10,t_B=5,t_C=20,t_D=15$. The last numbers, i.e., the timings of the pulses tell us that we have two sequential counterintuitive processes.

\begin{figure}
\begin{center}
\includegraphics[width=16cm]{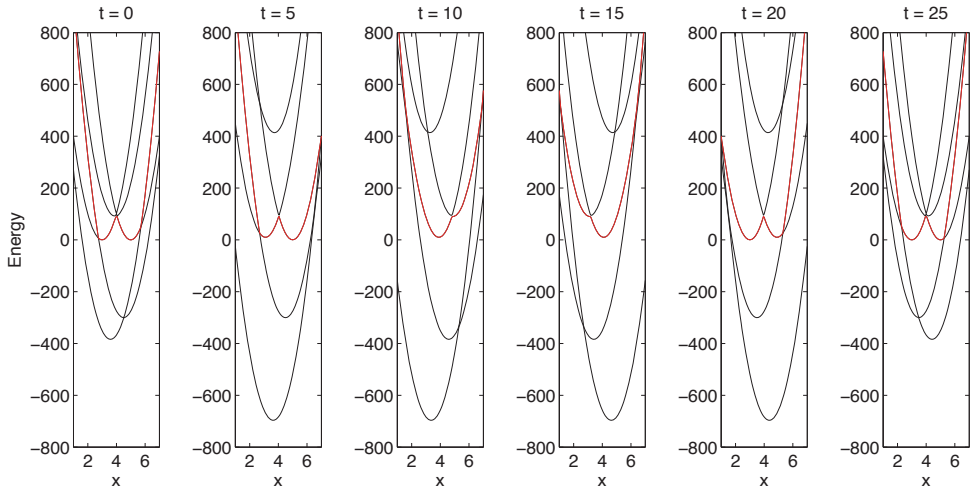}
\caption{APLIP process with five states as two sequential three-state APLIP processes: Eigenstates.}
\label{DoubleThreePot}
\end{center}
\end{figure}

To understand the APLIP process we can look at fig.~\ref{DoubleThreePot}. The double-well structure near zero-energy represents the relevant LIP, which smoothly (adiabatically) loses its double-well character, which is restored later but now the original ground state wave function has followed the LIP adiabatically and occupies the well on the right-hand side. Note that the main action appears to happen relatively fast between $t=10$ and $t=15$, but the relevant LIP dynamics has started much earlier and lasts much longer. The question now is that can we eliminate state nr. 3 completely?

\begin{figure}
\begin{center}
\begin{minipage}{160mm}
\subfigure[]{
\resizebox*{8cm}{!}{\includegraphics{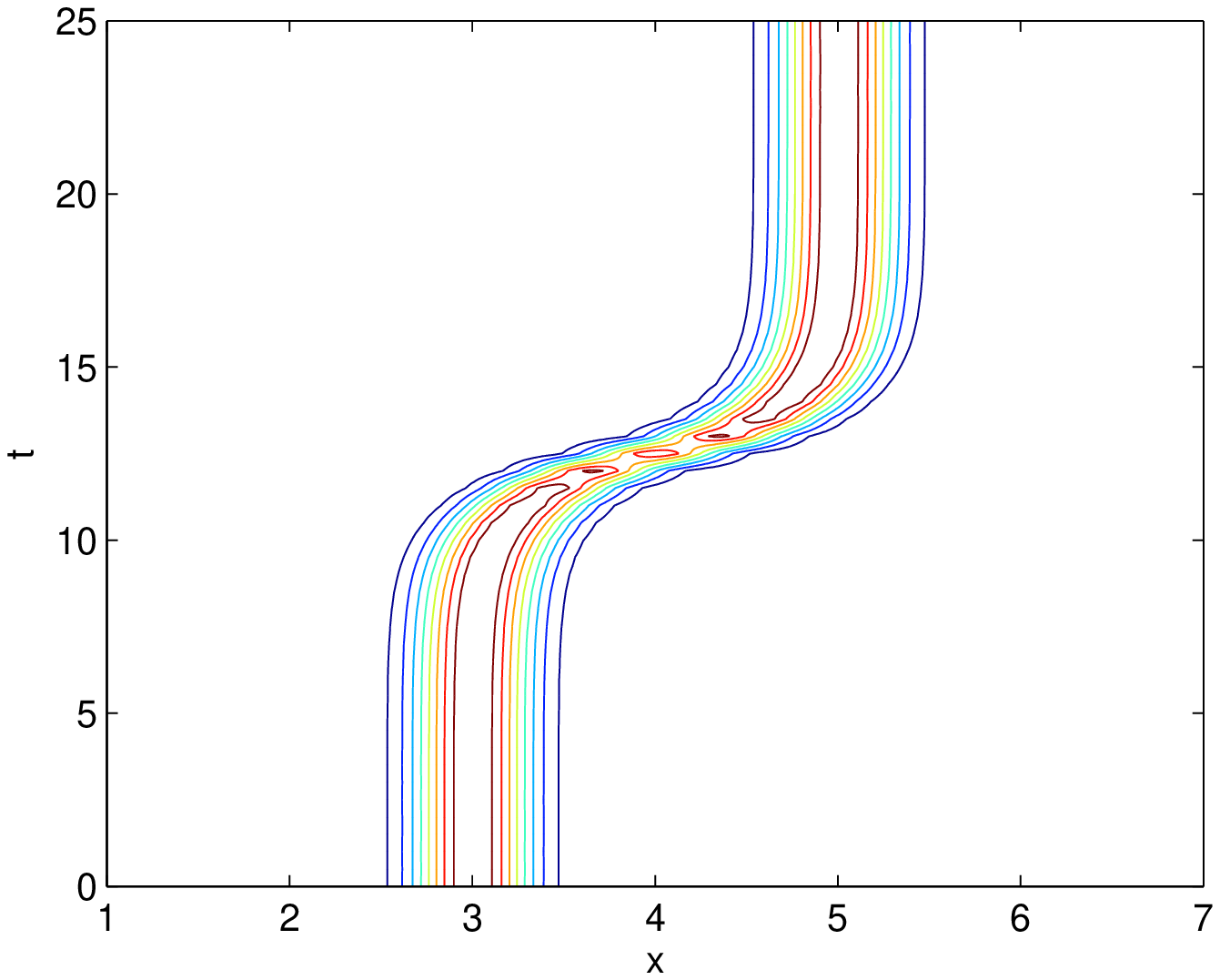}}}%
\subfigure[]{
\resizebox*{8cm}{!}{\includegraphics{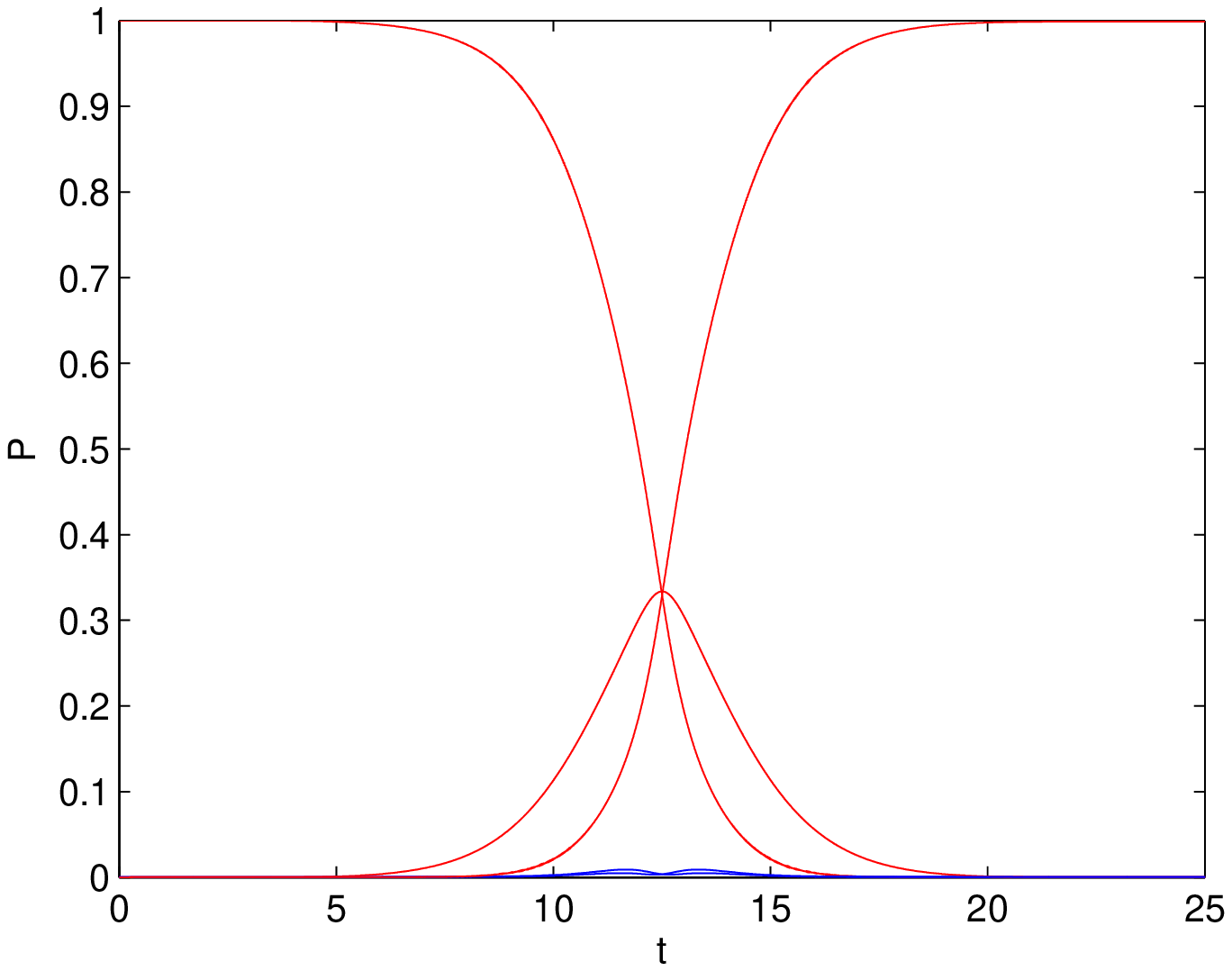}}}%
\caption{APLIP process with five states.}%
\label{FiveState}
\end{minipage}
\end{center}
\end{figure}

\begin{figure}
\begin{center}
\includegraphics[width=16cm]{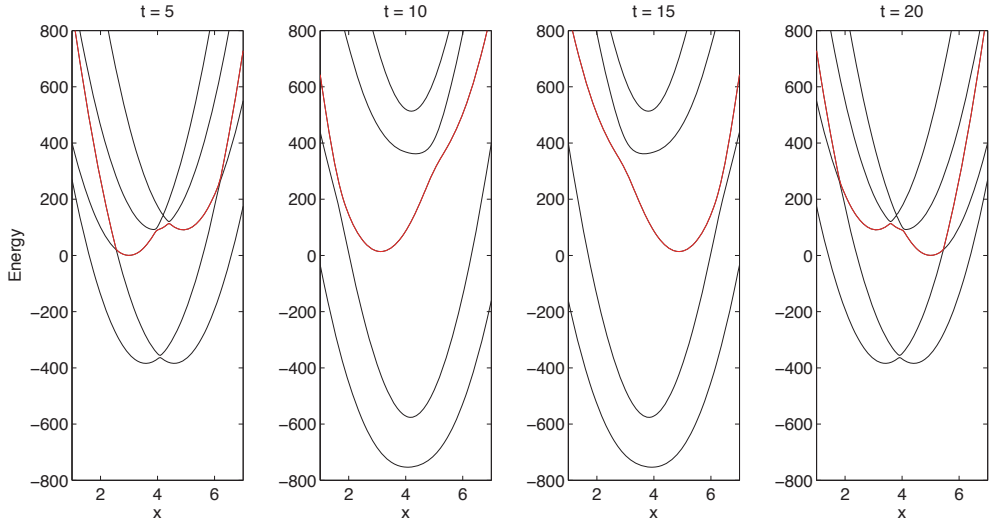}
\caption{APLIP process with five states: Eigenstate evolution.}
\label{FivePot}
\end{center}
\end{figure}

In fig.~\ref{FiveState}(a) one can see a seemingly swift transfer to the right, with full efficiency, and the data in (b) shows that the role of the intermediate state is strongly reduced. Here the scaled parameters are as previously except that $t_A=15,t_B=10,t_C=15,t_D=10$. So now the two counterintuitive sequences take place simultaneously. The eigenstate plots in fig.~\ref{FivePot} show now a process where one clearly has a deep LIP that takes the initial state to the final state. 

\begin{figure}
\begin{center}
\begin{minipage}{160mm}
\subfigure[]{
\resizebox*{8cm}{!}{\includegraphics{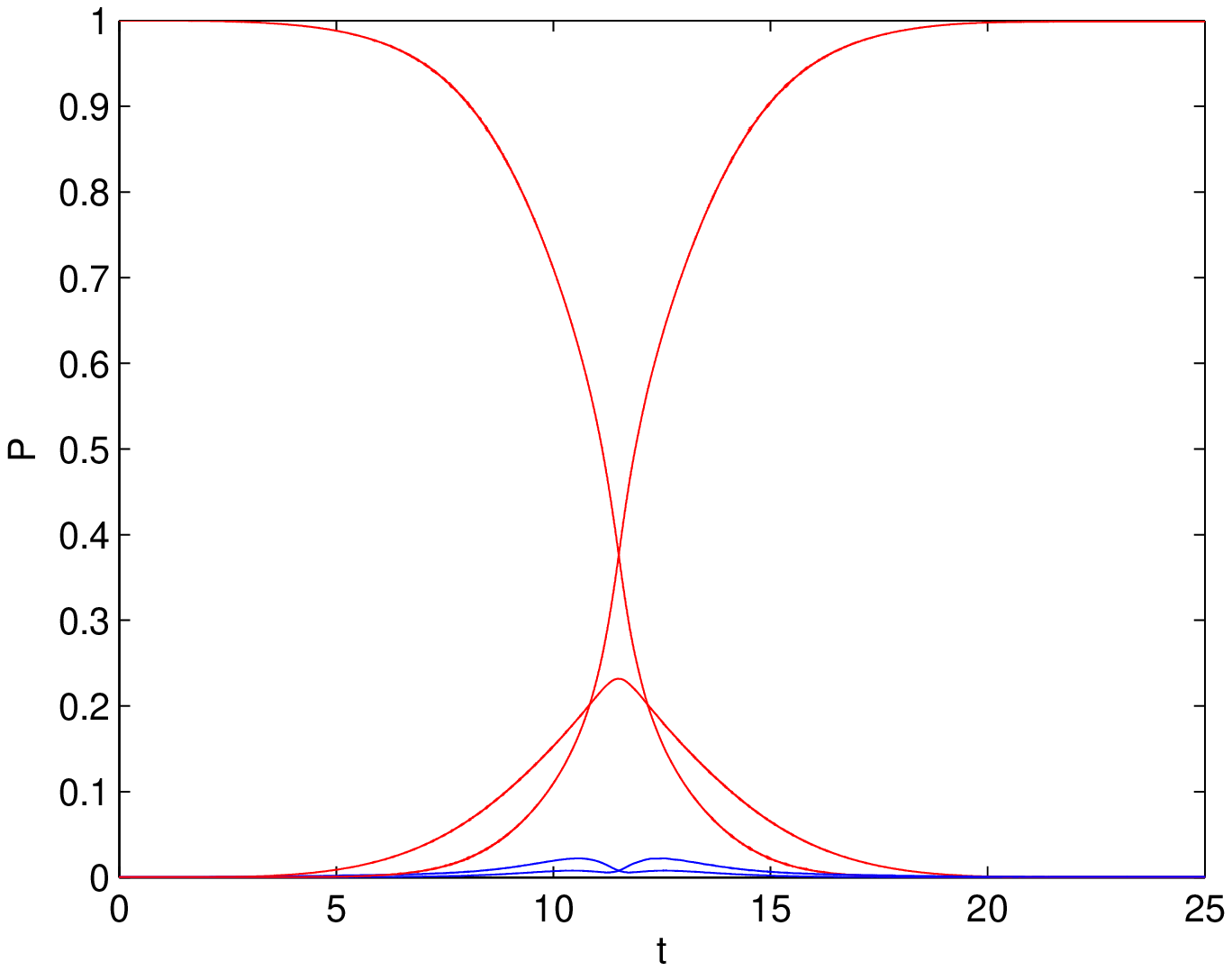}}}%
\subfigure[]{
\resizebox*{8cm}{!}{\includegraphics{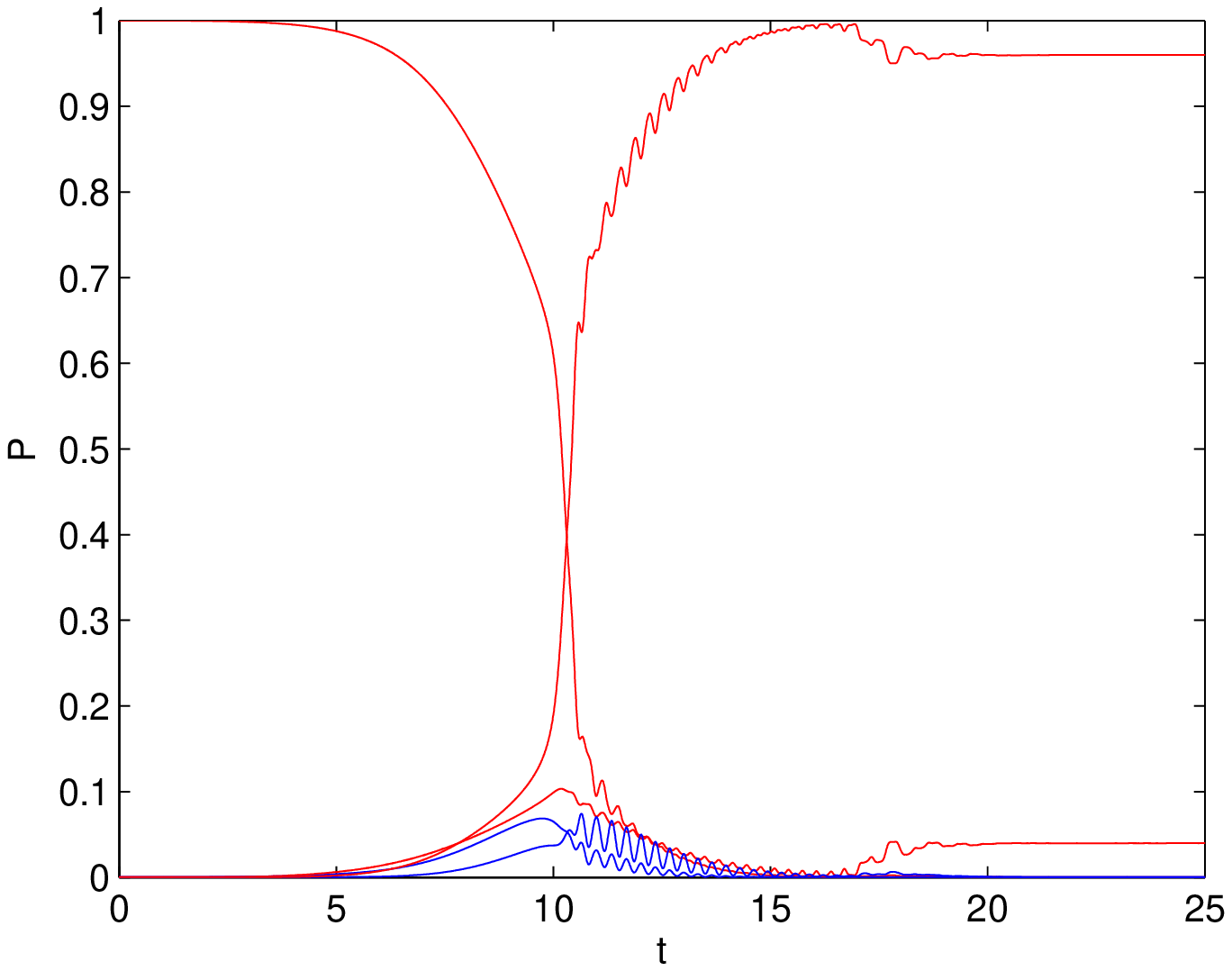}}}%
\caption{APLIP process with five states: Two examples.}%
\label{FivePopuTight}
\end{minipage}
\end{center}
\end{figure}

It remains to be asked whether we can take this even further e.g. by making the second set of two pulses even earlier. Figure~\ref{FivePopuTight} shows two examples with populations only. In (a) we have $t_A=15,t_B=10,t_C=13,t_D=8$ and $\Delta t_j=6$ for $j=A,B,C,D$, and in (b) $t_A=15,t_B=10,t_C=10,t_D=5$ with $\Delta t_j=6$. We can see that it is possible to reduce the role of the intermediate state even further, but eventually we start to lose the robustness of the process (in contour plots we would also see some oscillations in the final state, indicating that some vibrational states other than the ground state become excited).

\begin{figure}
\begin{center}
\begin{minipage}{160mm}
\subfigure[]{
\resizebox*{8cm}{!}{\includegraphics{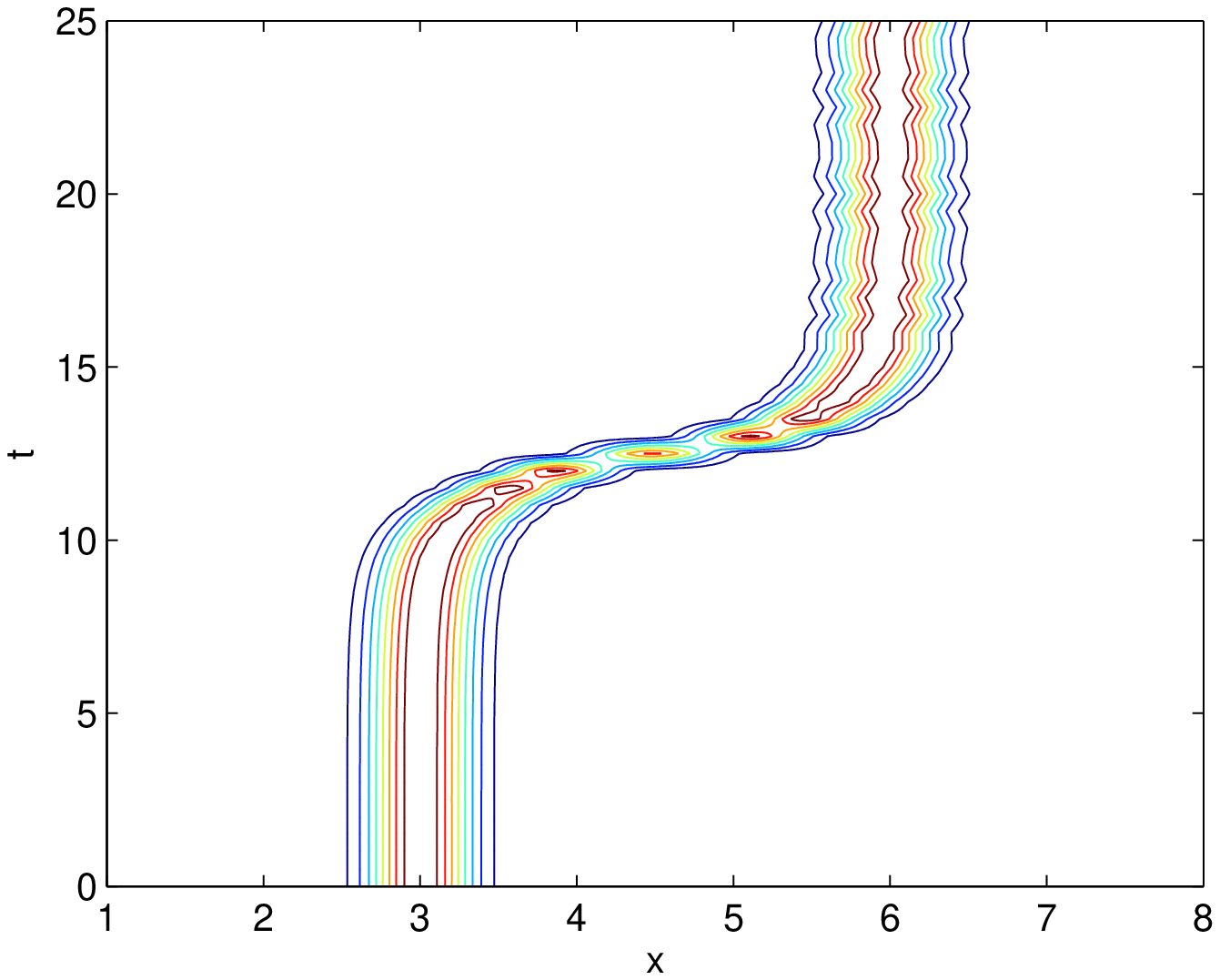}}}%
\subfigure[]{
\resizebox*{8cm}{!}{\includegraphics{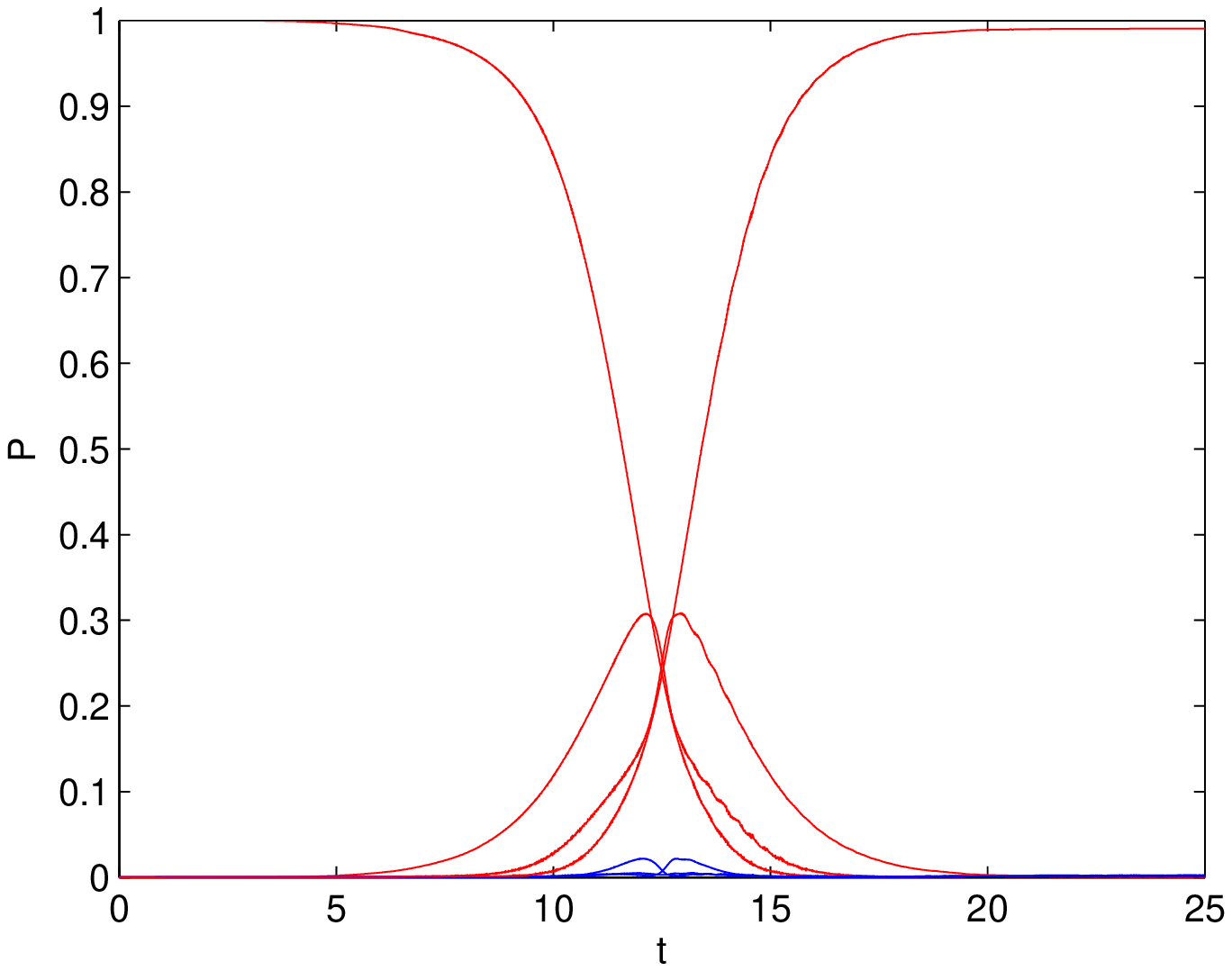}}}%
\caption{APLIP process with seven states.}%
\label{SevenState}
\end{minipage}
\end{center}
\end{figure}

\begin{figure}
\begin{center}
\includegraphics[width=16cm]{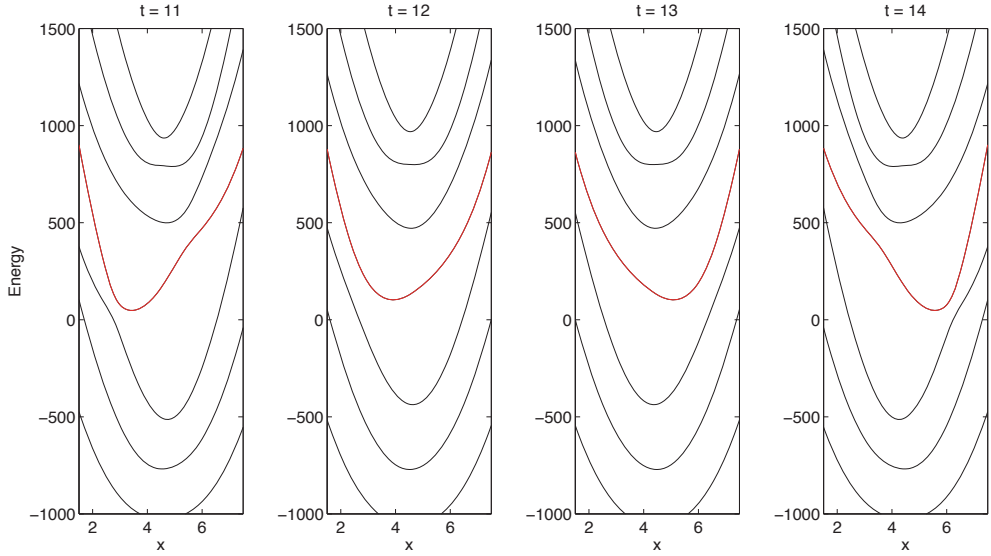}
\caption{APLIP process with seven states: Eigenstate evolution.}
\label{SevenPot}
\end{center}
\end{figure}

Getting ambitious, we now ask if APLIP is possible with seven states. We set the scaled parameters as $\Delta_1=\Delta_3=\Delta_5=\Delta_7=0$, $\Delta_2=\Delta_4=\Delta_6=-200$, $A_i=100$ for $i=1,..,7$, $\Delta t_j=5,B_0=700$ for $j=A,B,C,D,E,F$, $x_1=3.0,x_2=3.5,x_3=4.0, x_4=4.5,x_5=5.0, x_6=5.5,x_7=6.0$, and $t_A=15,t_B=10,t_C=15,t_D=10,t_E=10,t_F=15$. It means that, as in fig.~\ref{FiveState}, we have three counterintuitive processes exactly at the same time. The success of APLIP is almost excellent, as fig.~\ref{SevenState} demonstrates. The contour plot (a) shows some oscillations in the final state, but for populations (b) we see that even-numbered states have no population, and the two intermediate "ground" states 3 and 5 are only briefly occupied, at maximum with 0.3 of the total population (simultaneously at maximum with 0.4). The potential plot in fig.~\ref{SevenPot} demonstrates again the strong LIP character in dynamics. Note that in the three potential plots in this paper (figs.~\ref{DoubleThreePot}, \ref{FivePot} and \ref{SevenPot}) each plot looks at the crucial interval of change in increasing detail.

\begin{figure}
\begin{center}
\begin{minipage}{160mm}
\subfigure[]{
\resizebox*{8cm}{!}{\includegraphics{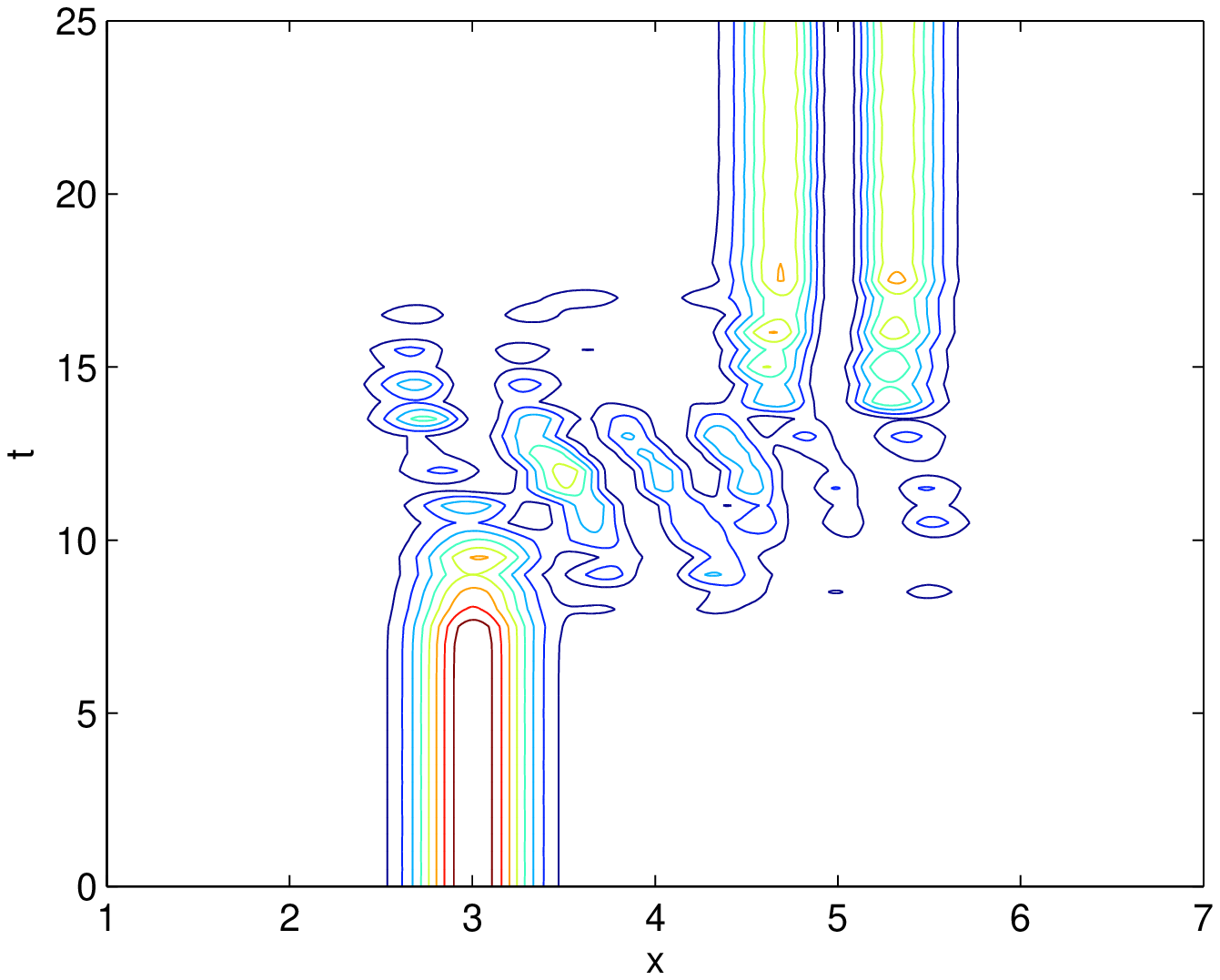}}}%
\subfigure[]{
\resizebox*{8cm}{!}{\includegraphics{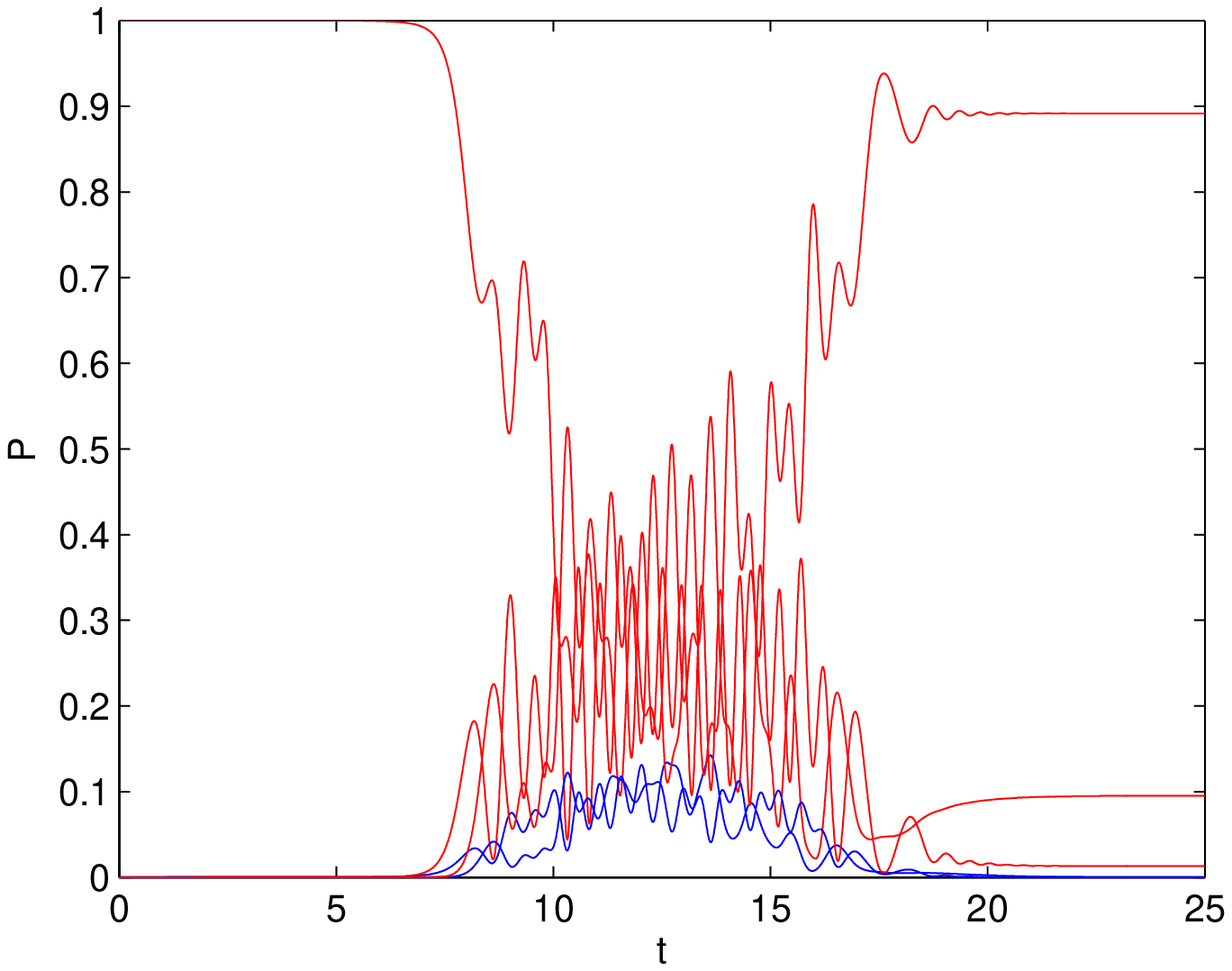}}}%
\caption{VibLIP process with five states.}%
\label{FiveVib}
\end{minipage}
\end{center}
\end{figure}

\begin{figure}
\begin{center}
\includegraphics[width=16cm]{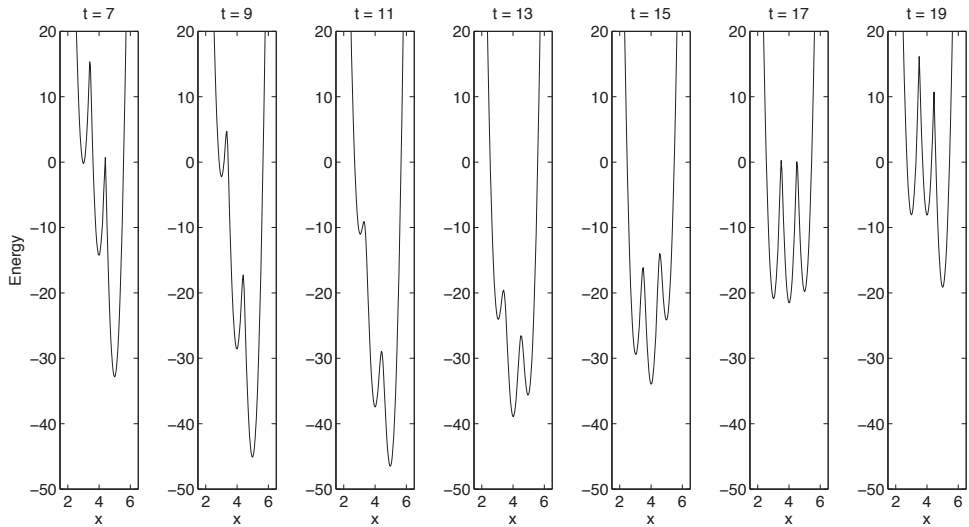}
\caption{APLIP process with five states: Eigenstate evolution.}
\label{FiveVibPot}
\end{center}
\end{figure}

Finally, we need to ask if the VibLIP process also works for the multistate case. Considering the wide variety of possible initial and final states (electronic and vibrational), I show only one example as a proof-of-principle. Finding the proper parameters is challenging since the process is not as robust as APLIP. We take again the five-state case with scaled parameters as $\Delta_1=\Delta_3=0$ but $\Delta_5=-19.1$, $\Delta_2=\Delta_4=850$, $A_i=100$ for $i=1,..,5$, $\Delta t_j=5,B_j=160$ for $j=A,B,C,D$, $x_1=3.0,x_2=3.5,x_3=4.0,x_4=4.5,x_5=5.0$, and $t_A=15,t_B=10,t_C=15,t_D=10$. The careful mapping of parameter space represented in ref.~\cite{Harkonen2006} is helpful in this even though it can not be directly mapped into the five-state case. In fig.~\ref{FiveVib} we see that we can indeed excite selectively the first excited vibrational state on the final state (a) with the efficiency of 0.9 (b). As with VibLIP in general, the intermediate populations of various states do not remain negligent. The potential plots in fig.~\ref{FiveVibPot} show that unlike APLIP, a successful VibLIP process requires a careful manipulation of the three-well LIP, which explains the missing robustness. Note also that for APLIP it is required that the even-numbered states have clearly negative $\Delta$'s whereas for VibLIP they are positive (for more discussion see ref.~\cite{Harkonen2006}).

\section{Conclusions}

From the numerical results it is clear that APLIP and VibLIP are clearly concepts that can be extended beyond the three-state problem into multiple states. The dynamics of the light-induced potentials are the key elements in understanding the process. Even in the case of two sequential three-state processes the full character of the five-state system is present in LIPs, as one can see in fig.~\ref{FivePot}. Such a multistate extension is not necessarily relevant in molecular physics with strong short pulses, but may find its uses in the physics of cold atoms~\cite{Weidemuller2009}. Already in ref.~\cite{Harkonen2006} it was shown that it is possible to combine APLIP and VibLIP approach with optical lattices. Perhaps a best target would be small samples of atoms (or even single atoms) above the quantum degeneracy. Another aspect is that both APLIP and VibLIP can serve as fictitious multistate systems when one looks for a time-dependent single state potential that can be applied to cold atoms, as discussed in ref.~\cite{Harkonen2006} as well. Finally, the applicability of APLIP and VibLIP for e.g. atoms that form a Bose-Einstein condensate~\cite{Pethick2008} is an intriguing topic, although preliminary studies indicate that the nonlinear term in the Gross-Pitaevskii equation is a source of problems.

\section*{Acknowledgements}

This work has been supported by the Academy of Finland grant 133682.

\bibliographystyle{apsrev4-1}

\end{document}